# Implicit Analysis of Perceptual Multimedia Experience Based on Physiological Response: A Review

Seong-Eun Moon, Jong-Seok Lee*, *Senior Member, IEEE*

*Abstract*—The exponential growth of popularity of multimedia has led needs for user-centric adaptive applications that manage multimedia content more effectively. Implicit analysis, which examines users' perceptual experience of multimedia by monitoring physiological or behavioral cues, has potential to satisfy such demands. Particularly, physiological signals categorized into cerebral physiological signals (electroencephalography, functional magnetic resonance imaging, and functional near-infrared spectroscopy) and peripheral physiological signals (heart rate, respiration, skin temperature, etc.) have recently received attention along with notable development of wearable physiological sensors. In this paper, we review existing studies on physiological signal analysis exploring perceptual experience of multimedia. Furthermore, we discuss current trends and challenges.

## I. Introduction

WITH the advances in the multimedia technology, the popularity of multimedia applications has been exponentially growing. As humans act as end-users of multimedia content, the ultimate goal of such applications is to satisfy the users by delivering appropriate content at the right moment in a proper way. Therefore, it is crucial to understand how users perceive multimedia content in order to provide user-centric services effectively. In a video streaming service, for example, it is necessary to understand the mechanism of visual quality perception in order to find the optimal operating mode that maximizes quality of experience (QoE) of delivered video content and, at the same time, minimizes the network resource usage [1]. As another example, monitoring the user's emotional state can be used for adaptive music recommendation to suggest music clips that match the current emotional state or help the user overcome negative emotion [2].

The perceptual experience of multimedia has diverse factors, including QoE, emotion, aesthetic satisfaction, preference, fatigue, attention, etc. In general, ways to recognize users' perceptual experience of multimedia can be broadly categorized as explicit and implicit approaches. The explicit approach, which has been traditionally adopted in numerous studies, employs human subjects, presents multimedia stimuli to them, and asks them to fill in a questionnaire. In some cases, detailed guidelines for conducting rigorous explicit evaluation tests with human subjects have been developed, e.g., standardized ITU recommendations for QoE assessment [3]–[5].

In contrast, the implicit approach does not require users' actions for answering questions or rating stimuli, but passively observes natural responses to given multimedia stimuli. Various channels convey cues of users' natural responses during multimedia consumption, from facial expression, gaze, and gesture, to physiological responses. In particular, implicit measurement using physiological signals, which this paper concentrates on[1], is of great interest due to the increasing popularity of wearable devices equipped with various physiological sensors. Physiological signals can be divided into two categories: brain activities and peripheral physiological activities. Cerebral signals measured by electroencephalography (EEG), magnetoencephalography (MEG), functional magnetic resonance imaging (fMRI), functional near-infrared spectroscopy (fNIRS), etc. carry comprehensive information about brain activities reflecting even the high-level cognitive process. On the other hand, peripheral physiological signals, i.e., galvanic skin response (GSR), skin temperature, respiration, heart rate, etc., are induced by the activities of the peripheral nervous system.

Although the explicit approach provides the most accurate and reliable results, the implicit approach has also received a significant amount of attention recently due to several advantages. First, while the explicit approach is mainly for off-line analysis, the implicit approach can be used for real-time monitoring of users' responses after training using ground truth data obtained from explicit methods. Second, the explicit approach may be prone to bias occurring in the process that users recall, conceptualize, and verbalize their experience, which is alleviated in the implicit approach.

There are a number of applications that can benefit from implicit analysis of perceptual multimedia experience, some of which are illustrated below (Fig. 1):

- QoE-aware content delivery: QoE of the multimedia content being delivered is monitored and used to adaptively adjust compression and network parameters to keep the perceptual quality of the delivered content acceptable and economize network and computing resources [8].
- Personalized content recommendation: By collecting implicitly monitored preference data over a period, the taste of the user is identified, based on which new content matching the taste is recommended [9].
- Content filtering: If a particular video clip has been recognized as containing inappropriate content (e.g., violent, horror, or pornographic) by analyzing some users' physiological responses to the clip, a content filtering engine prevents it from being shown to other particular viewers (e.g., children or minors).
- Implicit tagging: Users' responses to particular content can be used to automatically generate metadata such as tags (e.g., emotional tags, quality tags, etc.). It is well known that a large portion of multimedia data available online does not have appropriate metadata, which makes efficient content search and retrieval difficult. Thus, automatic metadata generation can be a powerful alternative to manual tagging [10], [11].
- Implicit control for disabled people: As the brain-computer interface (BCI) technology is beneficial for disabled people, techniques of implicit analysis of multimedia experience will be particularly useful for providing multimedia services to disabled users, e.g., image search and retrieval [12].
- Virtual reality (VR) and video games: Real-time monitoring of physiological responses enables a hand-free control of characters in VR and video games [13] and detection of particular user states such as fatigue [14], difficulty [15], and immersion [16]. Furthermore, treatments of mental disorders (autism, schizophrenia, attention deficit hyperactivity disorder, etc.) or age-related deficits through VR and video games can be implemented by using the physiological feedback or BCI technology [17], [18].
- Neuroscientific research: Understanding the relationship between physiological responses and multimedia stimuli is also interesting in the neuroscientific research. Relevant topics include identification of physiological channels appropriate for observing multimedia experience, understanding content features invoking particular physiological responses, etc.

---
[1]Overviews of implicit measurement techniques using behavioral cues can be found in [6], [7].

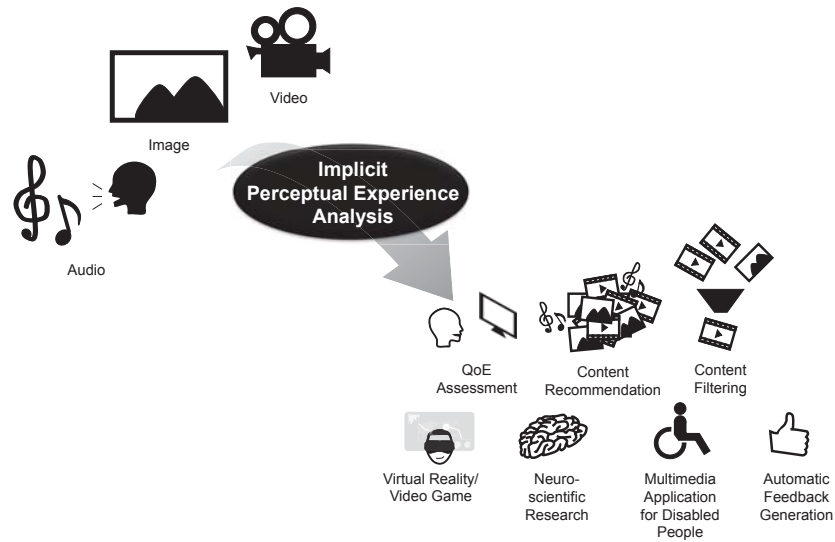

Fig. 1. Applications of physiological signal analysis of multimedia experience.

Fig. 2 shows a general system architecture for implicit analysis of perceptual experience of multimedia based on physiological signals. First, signals are acquired by sensors, for which pre-processing is applied to remove unwanted artifacts (e.g., blink artifact [19], muscle noise [20], etc.).

Then, representative features are extracted and used as inputs of a model that has been trained using machine learning techniques. The output of the model may be either class labels (e.g., positive/negative emotion, satisfied/unsatisfied, excellent/good/fair/poor/bad quality, etc.) or continuous values (e.g., valence or arousal values for emotion, quality ratings, etc.), depending on the problem dealt with.

The model part in Fig. 2 typically follows the general approach of machine learning, i.e., generic models such as neural networks or support vector machines (SVMs) can be used without significant modification. More critical issues in implicit analysis of multimedia experience are how to choose appropriate physiological signals, extract relevant features, and analyze the relationship between signal patterns and particular perceptual responses, which are focused in this paper.

In contrast to the previous reviews that consider only single perceptual factors (emotion [21], fatigue [22], and QoE [23]), we deal with several perceptual factors to provide a comprehensive view of the perception under the consideration that perceptual factors are often closely related with each other.

The remainder of the paper is organized as follows. Channels of the physiological signal measurement are briefly summarized in Section II. Then, Section III reviews recent studies of implicit perceptual experience of multimedia content with respect to the perceptual factors. Current trends and challenges are discussed in Section IV. Finally, concluding remarks are given in Section V.

II. PHYSIOLOGICAL SIGNALS FOR IMPLICIT ANALYSIS

In this section, we introduce various channels used to measure physiological signals and their representative features that can be used for implicit analysis of multimedia experience. Table I compares the characteristics of the physiological channels in terms of relevance to perceptual factors, temporal resolution, spatial resolution, and portability.

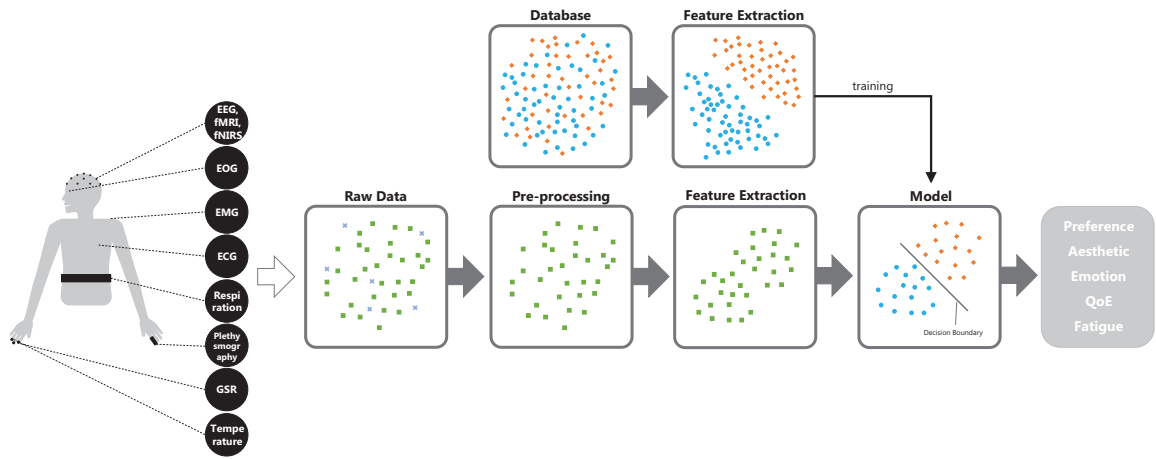

Fig. 2. General system of perceptual experience analysis using physiological signals.

TABLE I
COMPARISON OF PHYSIOLOGICAL SIGNAL CHANNELS

| Channel | Description | Temporal resolution | Spatial resolution | Portability | Factor |
|---|---|---|---|---|---|
| EEG | Measuring electrical activities on the scalp | ○ | × | △ | Emotion [24]–[38], QoE [39]–[50], fatigue [51] |
| MEG | Measuring magnetic fields induced by neuronal currents | ○ | △ | × | Emotion [52] |
| fMRI | Measuring cerebral blood flow by using the blood-oxygen-level dependent contrast | × | ○ | × | Emotion [53], [54], QoE [55], [56], aesthetics [57]–[59], fatigue [56], [60] |
| fNIRS | Measuring cerebral blood flow optically by using near-infrared light | △ | ○ | △ | QoE [61] |
| GSR | Measuring electric resistance that is decreased by secretion of the sweat | × | - | ○ | Emotion [27], [36]–[38], [62]–[64], QoE [48], aesthetics [65] |
| ECG, Plethysmograph | Monitoring heart rate that is increased when the sympathetic nerve is activated | × | - | ○ | Emotion [36]–[38], [52], [62]–[64], QoE [47], [48], [50], fatigue [66] |
| Respiration | Being slowed down in relaxation and irregular with negative emotions | × | - | ○ | Emotion [28], [36]–[38], [62]–[64], QoE [47], [48], [50] |
| Skin temperature | Being increased when the sympathetic nerve is activated | × | - | ○ | Emotion [36], [38], [64], [67], QoE [48] |
| EOG | Measuring ocular movement to obtain eye blinking signals | ○ | - | △ | Emotion [38], [52], [67] |
| EMG | Measuring muscular movement to detect facial expression or mental stress | ○ | - | △ | Emotion [38], [52], [62], [64], [67] |

○: Good, △: Fair, ×: Poor, -: Not applicable

## A. EEG

EEG monitors electrical activities induced by ionic flows within neurons through electrodes connected to the scalp, which enables to measure the brain activity with high temporal resolution. The international 10-20 system (Fig. 3) is typically employed to determine the locations of the electrodes. EEG has a long history in cerebral signal measurement; the first human study of EEG was conducted in 1924 [68]. EEG has excellent temporal resolution but relatively poor spatial resolution due to the volume conductance effect [69].

EEG signals can be represented by various types of features, such as event-related potentials, powers, and connectivity.

*1) Event-related potential (ERP):* An ERP indicates the change in the time series of an EEG signal as the direct result of a certain sensory, motor, or cognitive event. The event induces changes of phases and amplitudes of EEG signals, which typically occurs in a few hundred milliseconds before and after the event. Commonly, changes that occur within 100 ms after an event are influenced by the sensory inputs, whereas later changes reflect cognitive processes [70]. One of the most well-known ERP features is P300 [71], which is a positive peak appearing approximately 300 ms after an event. It was discovered in 1965 [72] and has been popularly employed for several neuroscientific researches [73], [74], particularly for selective attention [75] and information processing [76].

*2) Power-based features:* Another well-known type of features of EEG is the signal power of each electrode channel, which can measure the level of activeness of the neurons at a scalp region. In comparison to ERPs that mainly regard instant responses to certain events, power-based features account for responses aggregated over a time period. Usually, powers in different frequency bands, such as alpha (8-13 Hz), beta (14-30 Hz), gamma (31-50 Hz), delta (1-3 Hz), and theta (4-7 Hz), are separately extracted from EEG signals. Generally, high amplitude and low frequency signals are observed in calm states, and low amplitude and high frequency signals are observed in alert states [77]. Depending on the research purpose, the ranges of these frequency bands may differ or be divided into narrower subbands.

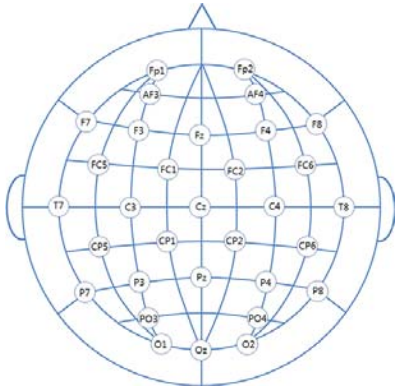

Fig. 3. International 10-20 system describing the locations of EEG electrodes. Each channel is identified by a letter-number combination. The letter indicates the region of the brain (i.e., F: frontal; Fp: frontal pole; C: central; P: parietal; O: occipital; T: temporal; AF: anterior-frontal). An odd or even number means the left or right side of the brain, and the midline is marked by letter z.

*3) Connectivity-based features:* While the aforementioned ERP and power-based features treat different regions of the brain separately, connectivity analysis concentrates on revealing how different brain regions are related. Brain connectivity is explained in three ways: structural connectivity, functional connectivity, and effective connectivity [78]. Structural connectivity concerns how different brain regions are anatomically connected. Functional connectivity measures (undirected) statistical dependence between separate brain regions. Finally, effective connectivity describes (directed) causal relationships between brain regions. What is called connectivity in the implicit analysis typically refers the functional and effective connectivity. Brain networks are described by vertices connected by edges. Usually, the vertices represent brain regions, and the edges represent connections or, when it is weighted, strength of connections. The strength of connections can be measured by various connectivity features such as correlation, phase synchronization, mutual information, and Granger causality [78]. Additionally, measures describing network characteristics can be derived from the connectivity features [79]. For example, the degree refers to the number of edges connected to an individual node, and its average is used as a measure of network density. The global efficiency, which is a measure of the level of integration in brain networks, is computed as an average of inversed shortest path lengths. Connectivity-based features have been only recently introduced to implicit analysis of multimedia experience, particularly for emotion [32], [33].

### B. MEG

MEG records the magnetic fields induced by electrical activities in the brain, which is also the source of the EEG measurement. MEG has a relatively long history; the first human study of MEG was performed in 1968 [80]. Although MEG and EEG measure the same cerebral activities, they have different characteristics. First, MEG costs higher and requires more space-consuming equipment than EEG. Furthermore, MEG has finer spatial resolution than EEG, but less sensitive to the electrical activities in the brain than EEG.

Measured MEG signals can be represented by various types of features including those described previously in Section II-A. Moreover, a mean value of the MEG signal over a certain time period can be utilized as an effective indicator of the activation of a brain region because of its fine spatial resolution [81].

### C. fMRI

fMRI identifies deoxy-hemoglobin in cerebral blood flow whose changes are associated with brain activities. The principle of fMRI was invented in 1990 [82], and the first human study was conducted in 1992 [83]. The fMRI method is extremely expensive and space-consuming. In addition, fMRI requires subjects to lie down in a magnet bore, which is usually more unusual and uncomfortable than EEG and fNIRS. This environmental restriction of fMRI also sets limits in multimedia consumption. As conventional displays cannot be used in the magnet bore, visual content needs to be projected by external devices usually having poor spatial resolution and color saturation or displayed on special devices that require additional cost. Also, noise of the fMRI scanner restricts the consumption of auditory content. Furthermore, fMRI cannot be performed on subjects who have metallic implants in their bodies. However, it has an advantage that it can produce brain images with outstanding spatial resolution. This allows to analyze the resulting images as activation of specific brain regions and, furthermore, to make connectivity-based analysis particularly effective.

### D. fNIRS

fNIRS is a relatively new technique for brain activity analysis. The first fNIRS-based analysis of human subjects was conducted in 1993 [84]. fNIRS identifies the hemoglobin status in blood flow as in fMRI, but uses near-infrared light instead, which allows only surficial measurement of neural activities. Similarly to fMRI, fNIRS has higher spatial resolution and lower temporal resolution than EEG. In addition, fNIRS costs less than fMRI and more robust to movement artifacts than EEG. Due to these advantages, fNIRS recently receives attention as a promising solution for cerebral signal measurement. Similarly to fMRI, results of fNIRS can be interpreted in terms of the activation of specific brain regions, and connectivity-based features can be also extracted.

### E. Peripheral physiological signals

Peripheral physiological signals reflect the state of the central and peripheral nervous systems such as arousal of the sympathetic nervous system. GSR, also called skin conductance, is one of the most popular channels for peripheral physiological signal monitoring. It typically measures the state of the sympathetic nervous system. That is, when sympathetic nerves are activated, sweat on the skin increases, which leads to an increase of skin conductance [85]. Heart rate, which is measured by plethysmography or electrocardiogram (ECG), is also a cue of the state of the sympathetic nervous system [86]. Respiration is generally slowed down in relaxation and becomes deep with negative emotions [87]. Skin temperature increases with the activation of the sympathetic nervous system. EOG and EMG measure ocular and muscular activities, respectively, by monitoring electrical changes that stem from movements. Typically, EOG is utilized to obtain the movement information of eyes, and EMG is adopted for detection of facial expression, head movement, and tension of muscle.

Signal statistics such as average, standard deviation, average of the first derivative, and standard deviation of the first derivative are usually employed as features of GSR, heart rate, and skin temperature. Temporal variability of heart rate and its energy (or the ratio between energies of low frequency and high frequency bands) can be also extracted as features. In the cases of EOG and EMG, their statistics are popularly used as features.

### F. Integration of brain and peripheral physiological signals

When the brain and peripheral physiological signals are compared in terms of classification or prediction performance of perceptual experience, the former usually outperforms the latter [50] [67]. Cerebral features tend to be more informative than features extracted from the peripheral signals in that the former contains whole information regarding the process of multimedia perception while only final outputs of the process (i.e., sympathetic responses) are observed in the latter. However, cerebral features are more likely to contain information irrelevant to the target factor than peripheral features such as the status of other part of body [88]. When they are measured simultaneously, they provide different views of the same perceptual experience, and thus synergic advantages can be expected via their integration.

In general, there are two approaches of multimodal integration for classification: feature fusion (or early integration) and decision fusion (or late integration). In the feature fusion approach, features of different modalities are combined into a composite feature vector under the assumption of perfect temporal synchronization, and then processed by a single classification system. In the decision fusion approach, classification is conducted for each modality independently and the outputs of the classifiers are combined to produce a final result. It has advantages over feature fusion in that it is relatively easy to use a weighting scheme to adjust relative amounts of contribution of the modalities and, moreover, asynchronous characteristics between the modalities can be considered easily [89].

Compromises of the two approaches are also possible. For instance, a concept of weak synchronization, called phase-amplitude coupling, was introduced to integrate cerebral and peripheral features (i.e., EEG and GSR) in [27]. Phase-amplitude coupling is a measure of the interaction between two signals having different frequencies, i.e., modulation of the amplitude of the higher frequency signal by the phase of the lower frequency signal. In [28], another approach to consider weak synchronization was proposed by measuring non-symmetric interdependence between frontal EEG and respiration.

## III. ANALYSIS OF PERCEPTUAL MULTIMEDIA EXPERIENCE

In this section, we review existing studies on implicit analysis of perceptual multimedia experience based on physiological signals. They are categorized according to the perceptual factors that are of interest in constructing practical multimedia services and applications: aesthetics, emotion, fatigue, QoE, and preference. In particular, representative recent studies are summarized in Table II.

### A. Emotion

Traditionally, numerous studies have explored various aspects of emotion, such as its nature [91]–[93], characteristics [94], expressions [95], [96], and influence on physical and mental health [97], [98]. Emotion appears relatively evident among other perceptual factors; body changes are often accompanied by emotion, for instance, smile, tear, sweating, heart pumping, shaking, flush, and so on. Furthermore, emotion influences individual and social behaviors [99]. A user's emotion is influenced primarily by semantic content of multimedia, however, high levels of quality and aesthetics may also facilitate induction of emotion. In real applications, it is crucial not only to detect emotion elicited by multimedia content but also to induce desired emotion to users.

There are many studies that tried to explain the relation between emotion and physiological signals. James and Lange [100] argued that emotions are feelings induced by physiological conditions. For instance, it is not that people cry because they feel sad, but rather people feel sad because they cry. However, some studies contradicted this theory, which is mainly supported by the fact that different emotions may induce similar reactions (e.g., fear and anger) [101]. In [102], the similarity of physiological responses induced by different emotions was explained by the influence of different situational context. According to this theory, an event is perceived as a stimulus that triggers a general autonomic arousal at first. Then, the physiological response is interpreted and labeled with a particular type of emotion through the contextual cognition of the event.

In fact, the difficulty of physiological analysis of emotion stems from the fact that physiological signals and emotion do not have one-to-one relationship but many-to-many relationship. That is, even though subjects feels the same emotion, their physiological reactions can vary according to the contextual environment, personality, and cultural background.

It is known that the amygdala, which is an almond-shaped structure located deep in the frontal portion of the temporal lobe, is closely related to emotional processes. Anatomical studies have shown that the amygdala is connected to various brain regions such as sensory systems, cognitive neocortical circuits, and the hypothalamus, and thus plays an important role in experience and expression of emotion [77]. In [53], negative emotions such as fear and sadness were induced by electrical stimulations of the right amygdala, and both positive and negative emotions were induced by electrical stimulations of the left amygdala. Relationship between the amygdala and emotions induced by music was explored using fMRI in [54]. In this study, it was found that dissonant music activates the amygdala and other regions including the hippocampus, parahippocampal gyrus, and temporal poles, which have strong functional and structural connection with the amygdala and are implicated in negative emotional processing.

TABLE II
REPRESENTATIVE STUDIES OF PHYSIOLOGICAL SIGNAL ANALYSIS

| Factor | Ref. | Media | Channel | Scheme | Results |
|---|---|---|---|---|---|
| Emotion | [54] | Music | fMRI | Comparison between pleasant (consonant) music and unpleasant (dissonant) music | Activations of amygdala, hippocampus, parahippocampal gyrus, and temporal poles |
|  | [29] | Music | EEG | Classification between joy, anger, sadness, and pleasure | Average accuracy of 85% for subject-dependent classification |
|  | [32] | Video | EEG | Classification of positive vs. negative valence | Synchronizations in the frontal and occipital sites with positive valence and synchronizations in the frontal lobe with negative valence |
|  | [62] | Music | Peripheral signals | Classification of high vs. low arousal and positive vs. negative valence | Average accuracies of 95% and 70% for subject-dependent and subject-independent classification |
|  | [38] | Music video | EEG & peripheral signals | Classification of high vs. low arousal, high vs. low valence, and like/dislike | F1 scores of 0.616, 0.647, and 0.618 for arousal, valence, and liking via decision fusion |
| QoE | [46] | 3D video | EEG | Correlation between QoE and EEG | Activation of the beta band in the right parietal lobe for high 3D QoE |
|  | [54] | Audiovisual stimuli | fMRI | Comparison between mismatched and well-matched pairs | Correlation between well-matched stimuli and activations of the inferior frontal gyrus, anterior superior insula, ventral striatum, etc. |
|  | [61] | Synthesized speech | fNIRS | Comparison between high and low quality synthesized speech | Increased brain activity in the orbitofrontal cortex for high quality synthesized speech |
|  | [48] | HDR video | EEG & peripheral signals | Classification of high vs. low dynamic range, good vs. poor contrast quality, and good vs. poor overall quality | Average accuracies of 80%, 80%, and 91% for subject-dependent classification and 55%, 57%, and 57% for subject-independent classification via decision fusion |
| Aesthetics | [57] | Image | fMRI | Comparison between gallery and computer contextual situations | Correlation between gallery context and activities of the medial orbitofrontal cortex and prefrontal cortex |
|  | [90] | Image | EEG | Comparison between more beautiful and less beautiful stimuli | Less beautiful stimuli elicit P200 components with higher amplitudes |
| Fatigue | [56] | 3D video | fMRI | Comparison between 2D and 3D TV watching conditions | Significant difference of brain activities between 2D and 3D |
|  | [66] | 3D video | Heart rate | Comparison between 2D and 3D TV watching conditions | Significantly increased heart rate and unstable autonomic state |

EEG has been actively utilized to investigate emotional states induced by multimedia content. Studies on ERP changes reflecting affective responses to images were reviewed in [24], which can be summarized as follows. ERP changes due to affective stimuli are more prominently observed in their amplitudes rather than latencies. ERPs from around 100 ms to several seconds after the events are modulated by affective stimuli, which reflects that emotion influences to several processing stages. In particular, the influence of valence tends to appear earlier than that of arousal. However, ERP changes by arousal have been observed more consistently than those by valence in several studies. In [25], it was attempted to exploit ERP changes due to viewing emotional images and obtained an accuracy of 81% for four-class classification (i.e., four quadrants of the valence-arousal plane).

The emotional responses to multimedia content have been frequently investigated using spectral power of EEG signals. Spectral power can be used directly as features or after conversion to other measures such as asymmetry index and differential entropy. In [26], relative power of the alpha, beta, gamma, delta, and theta frequency bands, defined by the spectral power of the frequency band divided by total spectral power of all frequency bands, was employed to build a system that recognizes emotions induced by videos (joy, neutral, anger, sad, and surprise). The asymmetry index, which refers to the level of asymmetry of neural activities in the left and right hemispheres of the brain, is regarded as an effective representation of emotion. In [29], for instance, spectral power changes and asymmetry indices of the alpha, beta, gamma, delta, and theta bands were employed for emotion classification (joy, anger, sadness, and pleasure) during music listening. A classification accuracy of 82% was obtained for a subject-dependent classification scheme, and it was shown that the features extracted from electrodes near the frontal and parietal lobes, which are known to be related to memory and sensory information processing, were particularly effective for classification. The differential entropy is an extension of Shannon entropy for a continuous random variable and used as a measure of the complexity. In [31], differential entropy was calculated from the spectral power distribution of each frequency band and used for classification of negative, neutral, and positive emotions induced by videos. As a result, an average classification accuracy of 87% was obtained, which is higher than that with power spectral features.

Connectivity features of EEG have been also employed for emotion analysis. In [32], the phase synchronization between EEG channels was examined under emotions of positive and negative valence (i.e., happiness and sadness) induced by videos. An overall increase of the phase synchronization was observed during emotional stimulation. In particular, happiness was associated with synchronization in a wide area of frontal and occipital sites, while sadness influenced only the frontal lobe. In [33], three connectivity measures, i.e., correlation, coherence, and phase synchronization between EEG channels were used for classification of three emotional states (positive, neutral, and negative) induced by emotion-eliciting film clips. Average accuracies of 54%, 61%, and 68% were obtained for the three types of features, respectively, which validated the effectiveness of the connectivity features for distinguishing emotional states.

Fusion of connectivity-based features and spectral power features was also explored. In [34], features that concern properties of the brain network consisting of EEG channels, such as global and local efficiency, small-worldness, etc. obtained from the magnitude squared coherence between EEG channels, were employed for classification of affective states (arousal, valence, dominance, and liking). The connectivity features outperformed spectral power and asymmetry index features in terms of the classification accuracy by 5-6%, and their combination additionally increased the classification performance. Similar results were obtained in [35] by employing mutual information of cross-frequency coupling patterns as connectivity features.

The information included in peripheral physiological signals is also valuable for affective analysis because emotion induces more intense physiological responses in comparison to other perceptual factors. Therefore, emotional responses to multimedia stimuli are relatively easy to capture via peripheral signals. For example, in [38], average accuracies using peripheral signal features for classification of high vs. low valence and liking vs. disliking were higher than those using EEG features.

Therefore, various peripheral physiological channels have been utilized for emotion detection. In [62], emotional responses to music (high vs. low arousal and positive vs. negative valence) were examined via GSR, ECG, EMG, and respiration. The EMG electrodes were placed at the upper trapezius muscle to measure the stress level. Statistical features of skin conductance, EMG, and subband spectrums of ECG, heart rate variability (HRV), spectral entropy and spectra of HRV, respiratory rate, subband spectrum of respiration signals, and breathing rate variability (BRV) were extracted. As a result, average classification accuracies of 95% and 70% were obtained from subject-dependent and subject-independent classification schemes, respectively. It was revealed that skin conductance and tension of the upper trapezius muscle increase according to the increase of the level of arousal. In addition, it was found that the features extracted from ECG and respiration were effective for classification of the valence level. Similar approaches were attempted for analysis of affective responses to video clips in [63] and [64]. The former investigated the perception of fear and sadness induced by using ECG, GSR, and respiration, and the latter tried to regress arousal and valence by using EMG, GSR, skin temperature, respiration, and plethysmograph.

Fusion of cerebral and peripheral physiological signals also has been attempted. Any advantage of the feature fusion scheme was not observed in [36] and [37], which investigated emotional states induced by images by combining EEG and peripheral physiological signals. The former employed power of EEG subbands and statistical features of heart rate, GSR, blood pressure, respiration, and temperature as classification features, and the latter employed statistical features of GSR, temperature, blood pressure, respiration, and raw EEG signals. In [38] and [52], decision fusion approaches were applied for combining cerebral signals (EEG and MEG, respectively) and peripheral physiological signals (GSR, heart rate, respiration, and skin temperature, etc.). The results demonstrated modest improvement of classification performance by fusion over single modalities. Although automatic determination of fusion weights remained unresolved, the synergy of the two modalities was validated.

It is difficult to compare the above-mentioned studies and analyze the superiority or inferiority of different methods, because they employed different scenarios and experimental protocols, such as emotional dimension, type of media, database, dataset partition for training and testing, and performance measure. Fortunately, a recent public database containing physiological affective response data for music videos, called DEAP (Database for Emotion Analysis using Physiological signals) [38], has been employed by several studies, some of which can be compared. Table III summarizes the studies that used the EEG data in the DEAP database based on similar experimental strategies in terms of emotional dimensions (arousal, valence, and liking), task (binary classification), classification scheme (subject-dependent scenario, leave-one-out cross-validation), and performance measures (accuracy and F1-score). It can be observed that connectivity-based features (in [34], [105]) tend to show better performance than the others. However, the highest accuracy reported remains only 76% for valence classification [105], which is not a satisfactory level yet, and thus efforts for further improvement will be desirable.

### B. QoE

QoE is defined as "the overall acceptability of an application or service, as perceived subjectively by the end-user" [106] or "the degree of delight or annoyance of the user of an application or service" [107]. It is closely related to but more user-centric than the traditional quality of service (QoS) that is rather device-, infrastructure-, and signal-centric, such as signal-to-noise ratio (SNR), delay, packet loss rate, etc. QoE of multimedia content is not only influenced by QoS but also related to characteristics of the human sensory systems such as Weber's law, nonuniform auditory and visual sensitivity functions, and just noticeable difference (JND) [108]. Furthermore, contextual information of users, e.g., expectations and environments of users, also has a decisive effect on QoE [109].

Subjective tests as explicit analysis have been popularly used for measuring QoE, where a number of subjects are hired and asked to rate the quality of presented stimuli. Standardized recommendations provide guidelines regarding test environments, stimulus presentation protocol, outlier detection, rating scales, procedures of subjective data analysis, etc. (e.g., [3]–[5]).

There are a number of studies that investigated QoE of multimedia content using physiological signals. Below, we review them by distinguishing those examining instant QoE changes, overall QoE, and QoE of emerging types of multimedia.

*1) Instant QoE changes:* Instantaneous cerebral responses, such as ERPs and EEG power changes, have been used in

TABLE III
COMPARISON OF EXISTING STUDIES THAT IMPLEMENTED EEG-BASED AFFECTIVE IMPLICIT ANALYSIS USING THE DEAP DATABASE [38] BY USING SIMILAR EXPERIMENTAL PROTOCOLS (I.E., SUBJECT-DEPENDENT BINARY CLASSIFICATION AND LEAVE-ONE-OUT CROSS-VALIDATION)

| Ref.* | Features | Feature selection | Classifier | Performance** | | |
|---|---|---|---|---|---|---|
| | | | | Arousal | Valence | Liking |
| [38] | Spectral power, asymmetry index | Fisher's linear discriminant | Gaussian naive Bayes classifier | ACC: 0.62 F1: 0.58 | ACC: 0.58 F1: 0.56 | ACC: 0.55 F1: 0.50 |
| [103] | Dual-tree complex wavelet packet transform time-frequency features | Combination of singular vector decomposition, QR factorization with column pivoting, and F-ratio | SVM | ACC: 0.67 F1: 0.57 | ACC: 0.65 F1: 0.55 | ACC: 0.71 F1: 0.51 |
| [104] | Gaussian mixture model-based dimensionality reduction of spectral power features | - | SVM | ACC: 0.67 | ACC: 0.71 | ACC: 0.70 |
| [35] | Mutual information between inter-hemispheric spectro-temporal patterns | Minimum redundancy maximum relevance algorithm | SVM | ACC: 0.61 F1: 0.61 | ACC: 0.61 F1: 0.62 | ACC: 0.60 F1: 0.61 |
| [105] | Pearson correlation coefficient, phase coherence, mutual information | Fisher's linear discriminant | SVM | ACC: 0.74 | ACC: 0.76 | - |
| [34] | Global efficiency, local efficiency, small-worldness coefficient | Minimum redundancy maximum relevance algorithm | Relevance vector machine | ACC: 0.68 F1: 0.68 | ACC: 0.65 F1: 0.65 | ACC: 0.67 F1: 0.65 |

* [34], [35] used only 75% of data for leave-one-out cross-validation and the rest for feature selection.
** ACC: classification accuracy; F1: F1-score

several studies to evaluate the perceptual influence of quality changes and artifact appearances. In [39], it was observed that the P300 response was delayed with weak quality degradation of speech signals, and the amplitude of P300 increased with stronger quality degradation of the signals. It was also reported that the amplitude of P300 was correlated with the magnitude of the video quality change [40]. In [41], power-based EEG features were used for video QoE analysis. Five different types of artifacts in videos, i.e., popping on person, popping, blurring on person, blurring, and ghosting on person, were recognized by EEG signals. Power increases were observed over the electrodes corresponding to the primary visual cortex (PO4, PO3, Oz, O1, and O2 in the international 10-20 system) for all types of artifacts. In addition, an accuracy of 85% was obtained for detection of the presence of artifacts using SVM classifiers. As for images, the study in [42] revealed that the existence of JPEG compression artifacts changes the ERP signals of the occipital area (O1, Oz, and O2 electrodes in the international 10-20 system). These studies demonstrate that abrupt quality changes in images, audio, and videos can be successfully detected by observing brain activities.

*2) Overall QoE:* Assessing overall QoE of multimedia content using cerebral physiological signals has been also investigated. Two approaches have been used in literature. One is to monitor the immediate response at the onset of a stimulus, and the other is to observe time-aggregated responses through the entire duration of stimulus presentation.

Studies taking the former approach showed that the amplitude of P300 is correlated with the reverberation time of speech [43] and the distortion level of videos [44], and inversely correlated with quality of synthesized speech [45]. It was consistently observed in these studies that lower quality or larger quality degradation of audios and videos induced more neural activities, which indicates that more distorted stimuli require more strenuous cognitive processing activities.

The following studies conducted analysis of time-aggregated responses. The cross-modal perception of audiovisual stimuli was investigated in [55] by using fMRI. The audio and visual stimuli were shown alone and in pairs of matched and mismatched conditions to subjects. It was observed that the left parahippocampal gyrus, left hippocampus of the medial temporal lobes, and lingual gyrus of the occipital cortex, which are known to be related to the memory and visual processing, were activated during the cross-modal condition compared to the unimodal condition. In addition, the mismatch effect of the audiovisual stimuli appeared in the prefrontal cortex, which is a storage of short-term memory and processes sensory inputs. Quality of synthesized speech signals was also investigated using fNIRS in [61]. The orbitofrontal cortex was more activated by high and medium quality signals in comparison to low quality signals.

Overall, it seems that brain regions activated or deactivated by stimuli differ depending on perceived quality, where the type of multimedia (i.e., speech, video, etc.) may also have an impact.

*3) QoE of emerging multimedia:* Furthermore, physiological signals have been used for analysis of perceptual experience of emerging multimedia technologies such as 3D, high dynamic range (HDR) imaging and 4K ultra high definition television (UHDTV). While the aforementioned studies focused on detecting QoE degradation of multimedia content, researches on QoE of emerging multimedia concentrate more on investigating perceptual consequences due to the effects intended by the technologies, such as immersiveness and sense of reality. Since these effects are induced after a user consumes given content for a certain duration, studies concerning them usually employ the aforementioned long-term signal analysis approach.

In [46], implicit monitoring of QoE of 2D and 3D videos was explored. The power spectral density of six frequency bands, namely, alpha, beta low (13-16 Hz), beta middle (17-20 Hz), beta high (21-29 Hz), gamma, and theta, was calculated, and its correlation with subjective ratings was investigated. As a result, it was shown that the frontal asymmetry patterns in the alpha band is related to the perceived quality. In [47], QoE of 4K UHD audiovisual content was evaluated in terms of the immersiveness level of the content. EEG and peripheral physiological signals (ECG and respiration) of subjects were recorded. From EEG, linear Granger causality features describing brain connectivity were calculated. HRV, respiratory rate, and frequency power of the respiration signal were extracted from the peripheral signals. When the two modalities were integrated via feature fusion for SVM-based three-class classification (high vs. middle vs. low immersiveness), low and high immersiveness levels were classified with accuracies of 61% and 94%, respectively. In [48], EEG and peripheral physiological signals (GSR, respiration, heart rate, and skin temperature) were employed to implicitly measure QoE of tone-mapped HDR videos in comparison to conventional low dynamic range videos. The frequency power of EEG was calculated from the theta, alpha, beta, and gamma frequency bands, and statistical features were obtained from the peripheral signals. From the results of the subject-dependent and subject-independent classification, it was shown that the power of the gamma frequency band, which is known to be related to sensory stimuli, was highly correlated with the perception of tone-mapped HDR videos. Surprisingly, classification accuracies obtained with peripheral features were higher than those with EEG features. Furthermore, decision fusion of the two modalities resulted in improved classification accuracies in the subject-dependent classification scheme. In [49], connectivity-based features were examined for the same task, where they enhanced classification performance significantly, particularly for subject-independent classification. In [50], EEG, ECG, and respiration were measured while subjects were watching 3D video stimuli and used for classification of the sensation of reality (high vs. low). Theta, alpha, beta, and gamma band frequency powers were extracted from EEG, and statistical features were calculated from the peripheral physiological signals. Additionally, a measure of signal complexity, called normalized length density index, was extracted from both types of signal. On average, a Matthews correlation coefficient (MCC) value of 0.16 was obtained using the peripheral physiological signals (ECG and respiration), which is only slightly higher than a chance level (i.e., zero), whereas the MCC value obtained using EEG was as high as 0.65.

### C. Aesthetics

Aesthetics is concerned with appreciation of beauty, including various connotations such as attractiveness or appealing. While it can be also defined for audio stimuli, aesthetics is typically considered for visual stimuli such as images and videos.

Immanuel Kant contended that the judgment of beauty is a judgment of taste based on a feeling of pleasure [110]. The pleasure in beauty is not a matter of being 'agreeable' or 'morally good'; it is desire-free (disinterested). Kant also argued that the judgment of beauty is both universal and subjective. Surely, the judgment of beauty is subjective as the judgment is based on personal taste. In addition, it is a process that can be sympathized and communicated.

Therefore, the perception of beauty is determined by the taste of visual factors (color, contrast, angle, orthogonality, symmetry, harmony, and so on) or their combinations. In photography, a few rules to enhance aesthetics of photographs have been empirically verified. For example, purer red, green, and blue appeal to viewers in nature photographs, and the rule of thirds, which states that a main object at the one third and two third lines catches viewers' eyes more than that at the exact center, is commonly accepted by photographers [111].

In [57], perception of aesthetics of images was investigated in the viewpoint of contextual situations. Abstract paintings were shown in two different contexts (gallery and computer) and subjects rated aesthetic scores, while the subjects' brain activities were measured via fMRI. As expected, images in the gallery context were rated significantly higher than those in the computer context. The ratings under the gallery context showed strong correlation with the activities of the medial orbitofrontal cortex and medial prefrontal cortex, while the ratings under the computer context showed no significant correlation with the brain signals. That is, aesthetic judgments of visual stimuli are related to the orbitofrontal cortex and prefrontal cortex, which are involved in the sensory integration and decision making process. Consistent results about the aesthetic perception of images were also obtained in [58] and [59] by using MEG and fMRI, respectively.

Generally speaking, it is difficult to recognize the aesthetic perceptual response by using peripheral physiological signals. Therefore, studies in that direction are rarely found. A recent study tried to identify aesthetic highlights by using GSR and obtained an average detection accuracy of 64%, but the detected aesthetic highlights may not correspond to scenes with high aesthetic quality but rather peaks of overall perceptual responses [65].

### D. Fatigue

Sometimes, multimedia consumption causes fatigue of users, which is apparently an undesirable effect. There are several causes of fatigue, such as too long a watching or hearing duration, a dim light condition in watching visual content, too short a distance between the screen and the eyes, too small a screen, etc. In particular, 3D fatigue, which is caused by 3D visual content, has been becoming a critical issue along with popularization of 3D multimedia [112]. Binocular parallax, which indicates difference of perceived images between the left and right eyes, gives a sense of depth to the human visual system. While different parts of stereoscopic images and videos have different amounts of binocular parallax, the accommodative distance (the distance to the point that the eyes must be focused) remains fixed as the distance to the display, unlike natural viewing where the varying binocular disparity is compensated by the eyes' vergence [22]. This conflict between vergence and accommodation is considered as a main cause of 3D visual fatigue.

The 3D visual fatigue was investigated via cerebral physiological signals in [51], [60], [56]. In [51], EEG responses in the beta frequency band significantly increased in the 3D environment, which received a significantly higher score in terms of visual fatigue in comparison to the case of 2D. It was observed via fMRI in [60] that the activation of the frontal eye field showed significant correlation with the amount of visual fatigue induced by the binocular disparties varied from 0° to 3°. In [56], watching 3DTV for a long duration (one hour) induced significant changes of fMRI measurements in the frontal eye field (Broadmann area 8), the visual cortices (Broadmann areas 17, 18, and 19), and other regions (Broadmann areas 32 and 40).

ECG was used for visual fatigue analysis in [66]. The heart rate increased significantly, and an increased HRV was observed during watching 3D videos. However, peripheral physiological signals are rarely employed for analysis of fatigue induced by multimedia consumption.

## IV. Trends and challenges

### A. Physiological signal sensors

In practice, the measuring of physiological signals requires significant efforts. For example, placing EEG electrodes on the scalp is not only time-consuming but also causes discomfort due to the interconnection via highly conductive gels to avoid artifacts induced by hair and enhance accuracies; in addition, the electrodes need to remain connected through wires to an external system composed of power supply, converters, control units, and so on. With the ever-increasing attention to wearable devices, however, more convenient and user-friendly devices have been released on the market. Commercial portable devices for EEG monitoring are now available, such as Emotiv [113], OpenBCI [114], NeuroSky [115], Mitsar portable EEG system [116], Avatar EEG [117], and Melomind [118]. In addition, devices that pursue enhanced potableness and convenience are under research and development (e.g., Ear-EEG) [119]. These allow portability of the devices, eliminate necessity of using conductive gels, and/or enable wireless data transmission. A few studies have tried to adopt such devices for the analysis of multimedia perceptual experience [120], [121].

Furthermore, smart bands and watches equipped with physiological signal sensors, such as Microsoft Band [122], Jawbone UP series [123], and Basis Peak [124], have been developed recently. These band type devices enable convenient real-time measurement of physiological signals in real life (typically, heart rate, GSR, and skin temperature).

In order to improve comfort and convenience, these devices tend to sacrifice performance in terms of spatial and temporal resolution, accuracy, robustness, etc. Therefore, it would be necessary to conduct benchmarking to validate the reliability of the portable consumer devices in comparison to the traditional ones used for lab-based measurement. Technological advances of wearable devices and rigorous validation of such devices would allow researchers to conduct studies of implicit multimedia experience analysis towards real-world applications easily. As an example, a recent study compared the performance of an open source EEG measurement platform called OpenBCI with a medical grade device priced around 25 times more expensive than OpenBCI [125]. The results indicated that while the performance of the medical grade equipment was slightly better in classification of a P300 speller and workload tasks, OpenBCI also showed comparable performance, and there were high temporal and spatial correlations between the signals acquired by the two systems.

### B. Deep learning approach

The deep learning approach, which has received significant attention recently as a powerful tool for learning of feature representation from data, has become a new trend of the machine learning-based physiological signal processing.

The performance of the deep learning approach on unsupervised feature extraction from raw physiological signals was verified in a few studies. In [126], it was shown that features learned from convolutional neural networks (CNNs) outperformed manually extracted features for affect classification of game players using GSR and blood volume pulse. In [127], a system based on deep belief networks (DBNs) was designed to recognize the levels of arousal, valence, and liking based on EOG and EMG, which was shown to be comparable to a Gaussian naive Bayes classifier with state-of-the-art expert-designed features. Nevertheless, further studies will be needed to examine potential of deep learning for performance improvement of analysis of perceptual multimedia experience and to investigate effective feature representation of physiological signals discovered via deep learning.

Another advantage of the deep learning approach is that it can easily handle multimodal data. (e.g., multimodal feature extraction using denoising auto-encoders [128], crossmodal learning and generalization using sparse restricted Boltzman machines [129]) In the future, similar techniques would be possible for fusion of cerebral and peripheral physiological signals, and furthermore, crossmodal learning among physiological signals and multimedia content.

### C. Preference analysis

Preference, which means liking or disliking of particular multimedia content, has been explained through the relationship with stimulus complexity in many studies [130]–[134]. The relationship between preference and stimulus complexity is commonly considered as an inverted U-curve, that is, neither too simple nor too complex stimuli are preferred. A recent study [134] revealed that there are different individual preference functions, and the inverted U-curve relationship appears as a combination of them.

The individual inclination of preference becomes critical when we consider the content of multimedia. For example, a person who likes animals may prefer a dog picture having poor quality to a building picture having good quality. Furthermore, individual experience, knowledge, and current needs and goals also have a significant influence to preference. Therefore, although the aforementioned aesthetics, emotion, fatigue, and QoE are all involved in determining preference, main factors that significantly influence to the decision of preference vary with the subject and situational context. That is, it is not straightforward to handle preference as a single perceptual factor to be analyzed. However, preference is often the final outcome of the perceptual processing of human subjects. In this sense, implicit analysis of preference is valuable for adaptive/personalized multimedia services. A preliminary study using NIRS demonstrated the potential of the implicit approach for detecting subjective preference [135]. Further studies will be desirable based on precise definition of preference and careful consideration of interrelated perceptual factors.

### D. Individual difference

The individual variance of perception is an obstacle to construct a generalized system for recognizing the perceptual status for multiple users. Even for the same stimulus, people perceive and respond differently depending on past experience and knowledge. It was demonstrated in [136] that subject-wise difference in fMRI data for the same stimuli is so large that even user identification is possible. As a result, the performance of recognizing perceptual experience in a subject-independent scheme is mostly poorer than that in a subject-dependent scheme, as reported in several studies (e.g., [62], [67], [137]). Although it is impossible to completely overcome such subjectivity, it will be desirable in the future to investigate ways of discovering physiological features common across different individuals.

### E. Environmental dependence

Even when the same multimedia stimulus is given to the same person, perception of the stimulus varies depending on the person's environment, e.g., mobile vs. desktop conditions, home vs. workplace, and alone vs. in crowds. For instance, in [138], it was shown that environmental noises significantly influenced to NIRS and peripheral physiological signals for music imagery tasks. However, environmental dependence has not been explored yet in the implicit approach, although explicit analysis has been conducted in several studies to reveal the influence of the environment (e.g., mobile phone vs. PDA vs. laptop for QoE [139], at home vs. on a vehicle vs. at school vs. at public space for QoE [140]). Investigating environmental dependence is becoming important as wearable physiological sensors enable measurement of users' perceptual experience under diverse situations.

### F. Open databases

In many research fields, it is crucial to have publicly available databases that enable benchmarking for fair comparison of different techniques and promote further related studies. The same applies to the research of implicit monitoring of physiological signal for perceptual multimedia analysis. However, there are only a few open databases that contain physiological signals and corresponding ground truth subjective labels in the context of multimedia experience, which are summarized in Table IV. Not only the number of available databases is small, but also most of them are for emotion analysis among various perceptual factors. Therefore, further efforts of the research community to create open databases targeting various factors of multimedia experience will be of great importance.

## V. CONCLUSION

We have reviewed approaches to monitoring perceptual experience of multimedia based on brain activities and peripheral physiological responses. The state-of-the-art studies were categorized, summarized, and compared with respect to the perceptual factor involved in multimedia experience and physiological modality, i.e., cerebral physiological signals and peripheral physiological signals. It was shown that cerebral signals are usually much informative in comparison to peripheral signals. However, this does not mean that the information obtained from peripheral signals is useless, particularly for analysis of emotional responses. Therefore, the fusion of cerebral and peripheral physiological signals is desirable, though the benefit of the fusion is not clearly evident yet. Furthermore, current trends and future challenges were discussed. When considering ever-increasing popularity of multimedia applications involving exponentially increasing volumes of multimedia data, we believe that implicit monitoring of multimedia experience will be of great value for adaptive multimedia services.

## ACKNOWLEDGMENT

This work was supported by the MSIP (Ministry of Science, ICT and Future Planning), Korea, under the "IT Consilience Creative Program" (IITP-R0346-16-1008) supervised by the IITP (Institute for Information & communications Technology Promotion).

## REFERENCES

[1] M. Cheon, S.-J. Kim, C.-B. Chae, and J.-S. Lee, "Quality assessment of mobile videos," in *Visual Signal Quality Assessment*. Springer International Publishing, 2015, pp. 99–127.

[2] J. J. Deng, C. H. C. Leung, A. Milani, and L. Chen, "Emotional states associated with music: classification, prediction of changes, and consideration in recommendation," *ACM Transactions on Interactive Intelligent Systems*, vol. 5, no. 1, pp. 4:1–36, 2015.

[3] ITU-R BT.500-13, "Methodology for the subjective assessment of the quality of television pictures." *International Telecommunication Union*, January 2012.

[4] ITU-R BT.2021, "Subjective methods for the assessment of stereoscopic 3DTV systems," *International Telecommunication Union*, February 2015.

[5] ITU-R BS.1284, "General methods for the subjective assessment of sound quality," *International Telecommunication Union*, December 2003.

[6] R. Reisenzein, M. Studtmann, and G. Horstmann, "Coherence between emotion and facial expression: evidence from laboratory experiments," *Emotion Review*, vol. 5, no. 1, pp. 16–23, 2013.

[7] A. T. Duchowski, "A breadth-first survey of eye-tracking applications," *Behavior Research Methods, Instruments, & Computers*, vol. 34, no. 4, pp. 455–470, 2002.

[8] J.-S. Lee, F. D. Simone, T. Ebrahimi, N. Ramzan, and E. Izquierdo, "Quality assessment of multidimensional video scalability," *IEEE Communications Magazine*, vol. 50, no. 4, pp. 38–46, 2012.

[9] C. Calcanis, V. Callaghan, M. Gardner, and M. Walker, "Towards end-user physiological profiling for video recommendation engines," in *Proceedings of the 4th International Conference on Intelligent Environments*, 2008, pp. 439–443.

[10] A. Yazdani, J.-S. Lee, and T. Ebrahimi, "Implicit emotional tagging of multimedia using EEG signals and brain computer interface," in *Proceedings of the 1st SIGMM Workshop on Social Media*, 2009, pp. 81–88.

[11] ——, "Toward emotional annotation of multimedia contents," in *Social Media Retrieval*. London: Springer, 2013, pp. 228–259.

TABLE IV
SUMMARY OF PUBLICLY AVAILABLE DATABASES

| | Channel | Media | Factor | Number of subjects |
|---|---|---|---|---|
| PsySyQX [141] | EEG and fNIRS | Speech | QoE and emotion | 21 |
| eNTERFACE2006 emotional database 1 [142] | EEG, fNIRS, facial videos, and peripheral signals (GSR, plethysmography, and respiration) | Image | Emotion | 5 |
| eNTERFACE2006 emotional database 2 [142] | fNIRS and facial videos | Image | Emotion | 16 |
| DEAP [38] | EEG and peripheral signals (GSR, blood volume, respiration, skin temperature, EMG, and EOG) | Music video | Emotion | 32 |
| MAHNOB-HCI [30] | EEG, eye gaze, audio, facial expression, and peripheral signals (ECG, GSR, respiration, and skin temperature) | Image and video | Emotion | 30 |
| SEED [31] | EEG | Chinese film | Emotion | 15 |
| DECAF [52] | MEG | Movie | Emotion | 42 |
| Yonsei-tHDRv [48] | EEG and peripheral signals (GSR, plethysmography, respiration, and skin temperature) | Tone-mapped HDR videos | QoE | 5 |

[12] A. D. Gerson, L. C. Parra, and P. Sajda, "Cortically coupled computer vision for rapid image search," *IEEE Transactions on Neural Systems and Rehabilitation Engineering*, vol. 14, no. 2, pp. 174–179, 2006.

[13] A. Lecuyer, F. Lotte, R. B. Reilly, R. Leeb, M. Hirose, and M. Slater, "Brain-computer interfaces, virtual reality, and videogames," *IEEE Computer*, vol. 41, no. 10, pp. 66–72, 2008.

[14] C. Zhao, M. Zhao, J. Liu, and C. Zheng, "Electroencephalogram and electrocardiograph assessment of mental fatigue in a driving simulator," *Accident Analysis & Prevention*, vol. 45, pp. 83–90, 2012.

[15] G. Chanel, C. Rebetez, M. Betrancourt, and T. Pun, "Emotion assessment from physiological signals for adaptation of game difficulty," *IEEE Transactions on Systems, Man, and Cybernetics-Part A: Systems and Humans*, vol. 41, no. 6, pp. 1052–1063, 2011.

[16] S. E. Kober and C. Neuper, "Using auditory event-related EEG potentials to assess presence in virtual reality," *International Journal of Human-Computer Studies*, vol. 70, no. 9, pp. 577–587, 2012.

[17] B. H. Cho, J. M. Lee, J. H. Ku, D. P. Jang, J. S. Kim, I. Y. Kim, J. H. Lee, and S. I. Kim, "Attention enhancement system using virtual reality and EEG biofeedback," in *Proceedings of the IEEE Virtual Reality*, 2002, pp. 156–163.

[18] J. A. Anguera, J. Boccanfuso, J. L. Rintoul, O. Al-Hashimi, F. Faraji, J. Janowich, E. Kong, Y. Larraburo, C. Rolle, E. Johnston, and A. Gazzaley, "Video game training enhances cognitive control in older adults," *Nature*, vol. 501, no. 7465, pp. 97–101, 2013.

[19] R. J. Croft and R. J. Barry, "Removal of ocular artifact from the EEG: a review," *Neurophysiologie Clinique/Clinical Neurophysiology*, vol. 30, no. 1, pp. 5–19, 2000.

[20] S. D. Muthukumaraswamy, "High-frequency brain activity and muscle artifacts in MEG/EEG: a review and recommendations," *Frontiers in Human Neuroscience*, vol. 7, p. 138, 2013.

[21] S. Jerritta, M. Murugappan, R. Nagarajan, and K. Wan, "Physiological signals based human emotion recognition: a review," in *Proceedings of the IEEE 7th International Colloquium on Signal Processing and its Applications*, 2011, pp. 410–415.

[22] M. Lambooij, M. Fortuin, I. Heynderickx, and W. Ijsselsteijn, "Visual discomfort and visual fatigue of stereoscopic displays: a review," *Journals of Imaging Science and Technology*, vol. 53, no. 3, pp. 30 201:1–14, 2009.

[23] J.-N. Antons, S. Arndt, R. Schleicher, and S. Moller, "Brain activity correlates of quality of experience," in *Quality of Experience*. Springer International Publishing, 2014, pp. 109–119.

[24] J. K. Olofsson, S. Nordin, H. Sequeira, and J. Polich, "Affective picture processing: an integrative review of ERP findings," *Biological Psychology*, vol. 77, no. 3, pp. 247–265, 2008.

[25] C. A. Frantzidis, C. Bratsas, C. L. Papadelis, E. Konstantinidis, C. Pappas, and P. D. Bamidis, "Toward emotion aware computing: an integrated approach using multichannel neurophysiological recordings and affective visual stimuli," *IEEE Transactions on Information Technology in Biomedicine*, vol. 14, no. 3, pp. 589–597, 2010.

[26] K. Kwang-Eun, H.-C. Yang, and K.-B. Sim, "Emotion recognition using EEG signals with relative power values and Bayesian network," *International Journal of Control, Automation and Systems*, vol. 7, no. 5, pp. 865–870, 2009.

[27] E. Kroupi, J.-M. Vesin, and T. Ebrahimi, "Implicit affective profiling of subjects based on physiological data coupling," *Brain-Computer Interfaces*, vol. 1, no. 2, pp. 85–98, 2014.

[28] ——, "Driver-response relationships between frontal EEG and respiration during affective audiovisual stimuli," in *Proceedings of the 35th Annual International Conference of the IEEE Engineering in Medicine and Biology Society*, 2013, pp. 2911–2914.

[29] Y.-P. Lin, C.-H. Wang, T.-P. Jung, and T.-L. Wu, "EEG-based emotion recognition in music listening," *IEEE Transactions on Biomedical Engineering*, vol. 57, no. 7, pp. 1798–1806, 2010.

[30] M. Soleymani, J. Lichtenauer, T. Pun, and M. Pantic, "A multimodal database for affect recognition and implicit tagging," *IEEE Transactions on Affective Computing*, vol. 3, no. 1, pp. 42–55, 2012.

[31] W.-L. Zheng and B.-L. Lu, "Investigating critical frequency bands and channels for EEG-based emotion recognition with deep neural networks," *IEEE Transactions on Autonomous Mental Development*, vol. 7, no. 3, pp. 162–175, 2015.

[32] T. Costa, E. Rognoni, and D. Galati, "EEG phase synchronization during emotional response to positive and negative film stimuli," *Neuroscience Letters*, vol. 406, no. 3, pp. 159–164, 2006.

[33] Y.-Y. Lee and S. Hsieh, "Classifying different emotional states by means of EEG-based functional connectivity patterns," *PloS One*, vol. 9, no. 4, p. e95415, 2014.

[34] R. Gupta, K. Laghari, and T. H. Falk, "Relevance vector classifier decision fusion and EEG graph-theoretic features for automatic affective state characterization," *Neurocomputing*, vol. 174, pp. 875–884, 2016.

[35] A. Clerico, R. Gupta, and T. H. Falk, "Mutual information between inter-hemispheric EEG spectro-temporal patterns: a new feature for automated affect recognition," in *Proceedings of the 7th International Conference on Neural Engineering*, 2015, pp. 914–917.

[36] G. Chanel, J. Kronegg, D. Grandjean, and T. Pun, "Emotion assessment: Arousal evaluation using EEG's and peripheral physiological signals," in *Proceedings of the International Workshop on Multimedia Content Representation, Classification and Security*, 2006, pp. 530–537.

[37] Z. Khalili and M. H. Moradi, "Emotion recognition system using brain and peripheral signals: using correlation dimension to improve the results of EEG," in *Proceedings of the International Joint Conference on Neural Networks*, 2009, pp. 1571–1575.

[38] S. Koelstra, C. Muhl, M. Soleymani, J.-S. Lee, A. Yazdani, T. Ebrahimi, T. Pun, A. Nijholt, and I. Patras, "DEAP: a database for emotion analysis; using physiological signals," *IEEE Transactions on Affective Computing*, vol. 3, no. 1, pp. 18–31, 2012.

[39] J.-N. Antons, R. Schleicher, S. Arndt, S. Moller, A. K. Porbadnigk, and G. Curio, "Analyzing speech quality perception using electroencephalography," *IEEE Journal of Selected Topics in Signal Processing*, vol. 6, no. 6, pp. 721–731, 2012.

[40] S. Scholler, S. Bosse, M. S. Treder, B. Blankertz, G. Curio, K. Muller, and T. Wiegand, "Toward a direct measure of video quality perception using EEG," *IEEE Transactions on Image Processing*, vol. 21, no. 5, pp. 2619–2629, 2012.

[41] M. Mustafa, S. Guthe, and M. Magnor, "Single-trial EEG classification

of artifacts in videos," *ACM Transactions on Applied Perception*, vol. 9, no. 3, pp. 12:1–15, 2012.

[42] L. Lindemann and M. Magnor, "Assessing the quality of compressed images using EEG," in *Proceedings of the International Conference on Image Processing*, 2011, pp. 3109–3112.

[43] J.-N. Antons, K. Laghari, S. Arndt, R. Schleicher, S. Moller, D. O'Shaughnessy, and T. H. Falk, "Cognitive, affective, and experience correlates of speech quality perception in complex listening conditions," in *Proceedings of the International Conference on Acoustics, Speech and Signal Processing*, 2013, pp. 3672–3676.

[44] S. Arndt, J. Antons, R. Schleicher, S. Moller, and G. Curio, "Using electroencephalography to measure perceived video quality," *IEEE Journal of Selected Topics in Signal Processing*, vol. 8, no. 3, pp. 366–376, 2014.

[45] S. Arndt, J.-N. Antons, R. Gupta, K. Laghari, R. Schleicher, S. Moller, and T. H. Falk, "Subjective quality ratings and physiological correlates of synthesized speech," in *Proceedings of the 5th International Workshop on Quality of Multimedia Experience*, 2013, pp. 152–157.

[46] E. Kroupi, P. Hanhart, J.-S. Lee, M. Rerabek, and T. Ebrahimi, "EEG correlates during video quality perception," in *Proceedings of the 22nd European Signal Processing Conference*, 2014, pp. 2135–2139.

[47] A.-F. Perrin, H. Xu, E. Kroupi, M. Rerabek, and T. Ebrahimi, "Multimodal dataset for assessment of quality of experience in immersive multimedia," in *Proceedings of the 23rd Annual ACM Conference on Multimedia*, 2015, pp. 1007–1010.

[48] S.-E. Moon and J.-S. Lee, "Perceptual experience analysis for tone-mapped HDR videos based on EEG and peripheral physiological signals," *IEEE Transactions on Autonomous Mental Development*, vol. 7, no. 3, pp. 236–247, 2015.

[49] ——, "EEG connectivity analysis in perception of tone-mapped high dynamic range videos," in *Proceedings of the 23rd ACM International Conference on Multimedia*, 2015, pp. 987–990.

[50] E. Kroupi, P. Hanhart, J.-S. Lee, M. Rerabek, and T. Ebrahimi, "Predicting subjective sensation of reality during multimedia consumption based on EEG and peripheral physiological signals," in *Proceedings of the IEEE International Conference on Multimedia and Expo*, 2014.

[51] Y.-J. Kim and E. C. Lee, "EEG based comparative measurement of visual fatigue caused by 2D and 3D displays," in *Proceedings of the Communications in Computer and Information Science (Part V)*, 2011, pp. 289–292.

[52] M. K. Abadi, R. Subramanian, S. M. Kia, P. Avesani, I. Patras, and N. Sebe, "DECAF: MEG-based multimodal database for decoding affective physiological responses," *IEEE Transactions on Affective Computing*, vol. 6, no. 3, pp. 209–222, 2015.

[53] L. Lanteaume, S. Khalfa, J. Regis, P. Marquis, P. Chauvel, and F. Bartolomei, "Emotion induction after direct intracerebral stimulations of human amygdala," *Cerebral Cortex*, vol. 17, no. 6, pp. 1307–1313, 2007.

[54] S. Koelsch, T. Fritz, D. Y. V. Cramon, K. Muller, and A. D. Friederici, "Investigating emotion with music: an fMRI study," *Human Brain Mapping*, vol. 27, no. 3, pp. 239–250, 2006.

[55] M. O. Belardinelli, C. Sestieri, R. D. Matteo, F. Delogu, C. D. Gratta, A. Ferretti, M. Caulo, A. Tartaro, and G. L. Romani, "Audio-visual crossmodal interactions in environmental perception: an fMRI investigation," *Cognitive Processing*, vol. 5, no. 3, pp. 167–174, 2004.

[56] C. Chen, J. Wang, K. Li, Y. Liu, and X. Chen, "Visual fatigue caused by watching 3DTV: an fMRI study," *Biomedical Engineering Online*, vol. 14 (Suppl 1), 2015.

[57] U. Kirk, M. Skov, O. Hulme, M. S. Christensen, and S. Zeki, "Modulation of aesthetic value by semantic context: an fMRI study," *Neuroimage*, vol. 44, no. 3, pp. 1125–1132, 2009.

[58] C. J. Cela-Conde, G. Marty, F. Maestu, T. Ortiz, E. Munar, A. Fernandez, and F. Quesney, "Activation of the prefrontal cortex in the human visual aesthetic perception," *Proceedings of the National Academy of Sciences of the United States of America*, vol. 101, no. 16, pp. 6321–6325, 2004.

[59] H. Kawabata and S. Zeki, "Neural correlates of beauty," *Journal of Neurophysiology*, vol. 91, no. 4, pp. 1699–1705, 2004.

[60] D. Kim, Y. J. Jung, E. Kim, Y. M. Ro, and H. Park, "Human brain response to visual fatigue caused by stereoscopic depth perception," in *Proceedings of the 17th International Conference on Digital Signal Processing*, 2011, pp. 1–5.

[61] R. Gupta, K. Laghari, S. Arndt, R. Schleicher, S. Moller, D. O'Shaughnessy, and T. H. Falk, "Using fNIRS to characterize human perception of TTS system quality, comprehension, and fluency: preliminary findings," in *Proceedings of the International Workshop on Perceptual Quality of Systems*, 2013, pp. 73–78.

[62] J. Kim and E. André, "Emotion recognition based on physiological changes in music listening," *IEEE Transactions on Pattern Analysis and Machine Intelligence*, vol. 30, no. 12, pp. 2067–2083, 2008.

[63] S. D. Kreibig, F. H. Wilhelm, W. T. Roth, and J. J. Gross, "Cardiovascular, electrodermal, and respiratory response patterns to fear-and sadness-inducing films," *Psychophysiology*, vol. 44, no. 5, pp. 787–806, 2007.

[64] M. Soleymani, G. Chanel, J. J. M. Kierkels, and T. Pun, "Affective characterization of movie scenes based on multimedia content analysis and user's physiological emotional responses," in *Proceedings of the 10th International Symposium on Multimedia*, 2008, pp. 228–235.

[65] T. Kostoulas, G. Chanel, M. Muszynski, P. Lombardo, and T. Pun, "Identifying aesthetic highlights in movies from clustering of physiological and behavioral signals," in *Proceedings of the 7th International Workshop on Quality of Multimedia Experience*, 2015, pp. 1–6.

[66] S. Park, M. J. Won, S. Mun, E. C. Lee, and M. Whang, "Does visual fatigue from 3D displays affect autonomic regulation and heart rhythm?" *International Journal of Psychophysiology*, vol. 92, no. 1, pp. 42–48, 2014.

[67] A. Yazdani, J.-S. Lee, J.-M. Vesin, and T. Ebrahimi, "Affect recognition based on physiological changes during the watching of music videos," *ACM Transactions on Interactive Intelligent Systems*, vol. 2, no. 1, pp. 7:1–25, 2012.

[68] L. F. Haas, "Hans Berger (1873-1941), Richard Caton (1842-1926), and electroencephalography," *Journal of Neurology, Neurosurgery & Psychiatry*, vol. 74, no. 1, pp. 9–9, 2003.

[69] S. P. van den Broek, F. Reinders, M. Donderwinkel, and M. J. Peters, "Volume conduction effects in EEG and MEG," *Electroencephalography and Clinical Neurophysiology*, vol. 106, no. 6, pp. 522–534, 1998.

[70] S. Sur and V. K. Sinha, "Event-related potential: an overview," *Industrial Psychiatry Journal*, vol. 18, no. 1, pp. 70–73, 2009.

[71] A. P. F. Key, G. O. Dove, and M. J. Maguire, "Linking brainwaves to the brain: an ERP primer," *Developmental Neuropsychology*, vol. 27, no. 2, pp. 183–215, 2005.

[72] S. Sutton, M. Braren, J. Zubin, and E. R. John, "Evoked-potential correlates of stimulus uncertainty," *Science*, vol. 150, no. 3700, pp. 1187–1188, 1965.

[73] S. H. Patel and P. N. Azzam, "Characterization of N200 and P300: selected studies of the event-related potential," *International Journal of Medical Science*, vol. 2, no. 4, pp. 147–154, 2005.

[74] J. Polich, "Updating P300: an integrative theory of P3a and P3b," *Clinical Neurophysiology*, vol. 118, no. 10, pp. 2128–2148, 2007.

[75] H. M. Gray, N. Ambady, W. T. Lowenthal, and P. Deldin, "P300 as an index of attention to self-relevant stimuli," *Journal of Experimental Social Psychology*, vol. 40, no. 2, pp. 216–224, 2004.

[76] C. C. Duncan-Johnson and E. Donchin, "The P300 component of the event-related brain potential as an index of information processing," *Biological Psychology*, vol. 14, pp. 1–52, 1982.

[77] M. F. Bear, B. Connors, and M. Paradiso, *Neuroscience: Exploring the Brain*. Williams & Wilkins, 2015.

[78] K. J. Friston, "Functional and effective connectivity: a review," *Brain Connectivity*, vol. 1, no. 1, pp. 13–36, 2011.

[79] M. Rubinov and O. Sporns, "Complex network measures of brain connectivity: uses and interpretations," *Neuroimage*, vol. 52, no. 3, pp. 1059–1069, 2010.

[80] D. Cohen, "Magnetoencephalography: evidence of magnetic fields produced by alpha-rhythm currents," *Science*, vol. 161, no. 3843, pp. 784–786, 1968.

[81] A. Hillebrand, K. D. Singh, I. E. Holliday, P. L. Furlong, and G. R. Barnes, "A new approach to neuroimaging with magnetoencephalography," *Human Brain Mapping*, vol. 25, no. 2, pp. 199–211, 2005.

[82] S. Ogawa, T. M. Lee, A. R. Kay, and D. W. Tank, "Brain magnetic resonance imaging with contrast dependent on blood oxygenation," *Proceedings of the National Academy of Sciences*, vol. 87, no. 24, pp. 9868–9872, 1990.

[83] S. Ogawa, D. W. Tank, R. Menon, J. M. Ellermann, S. G. Kim, H. Merkle, and K. Ugurbil, "Intrinsic signal changes accompanying sensory stimulation: functional brain mapping with magnetic resonance imaging," *Proceedings of the National Academy of Sciences*, vol. 89, no. 13, pp. 5951–5955, 1992.

[84] Y. Hoshi and M. Tamura, "Dynamic multichannel near-infrared optical imaging of human brain activity," *Journal of Applied Physiology*, vol. 75, no. 4, pp. 1842–1846, 1993.

[85] J. D. Montagu and E. M. Coles, "Mechanism and measurement of the galvanic skin response," *Psychological Bulletin*, vol. 65, no. 5, pp. 261–279, 1966.

[86] A. Schafer and J. Vagedes, "How accurate is pulse rate variability as an estimate of heart rate variability? A review on studies comparing photoplethysmographic technology with an electrocardiogram," *International Journal of Cardiology*, vol. 166, no. 1, pp. 15–29, 2013.

[87] F. A. Boiten, N. H. Frijda, and C. J. E. Wientjes, "Emotions and respiratory patterns: review and critical analysis," *International Journal of Psychophysiology*, vol. 17, no. 2, pp. 103–128, 1994.

[88] S. Sanei and J. A. Chambers, *EEG Signal Processing*. John wiley & Sons, 2013.

[89] J.-S. Lee and C. H. Park, "Robust audio-visual speech recognition based on late integration," *IEEE Transactions on Multimedia*, vol. 10, no. 5, pp. 767–779, 2008.

[90] X. Wang, Y. Huang, Q. Ma, and N. Li, "Event-reltaed potential P2 correlates of implicit aesthetic experience," *Neuroreport*, vol. 23, no. 14, pp. 862–866, 2012.

[91] N. H. Frijda, "The laws of emotion," *American Psychologist*, vol. 43, no. 5, pp. 349–358, 1988.

[92] M. Cabanac, "What is emotion?" *Behavioural Processes*, vol. 60, no. 2, pp. 69–83, 2002.

[93] P. Winkielman and K. C. Berridge, "Unconscious emotion," *Current Directions in Psychological Science*, vol. 13, no. 3, pp. 120–123, 2004.

[94] P. Ekman, W. V. Friesen, M. O'Sullivan, A. Chan, I. Diacoyanni-Tarlatzis, K. Heider, R. Krause, W. A. LeCompte, T. Pitcairn, P. E. Ricci-Bitti, K. Scherer, M. Tomita, and A. Tzavaras, "Universals and cultural differences in the judgments of facial expressions of emotion," *Journal of Personality and Social Psychology*, vol. 53, no. 4, pp. 712–717, 1987.

[95] D. Keltner and P. Ekman, *Facial Expression of Emotion*. Guilford, 2003.

[96] P. N. Juslin and P. Laukka, "Emotional expression in speech and music," *Annals New York Academy of Sciences*, vol. 1000, no. 1, pp. 279–282, 2003.

[97] L. S. Richman, L. Kubzansky, J. Maselko, I. Kawachi, P. Choo, and M. Bauer, "Positive emotion and health: going beyond the negative," *Health Psychology*, vol. 24, no. 4, pp. 422–429, 2005.

[98] M. Berking and P. Wupperman, "Emotion regulation and mental health: recent findings, current challenges, and future directions," *Current Opinion in Psychiatry*, vol. 25, no. 2, pp. 128–134, 2012.

[99] R. J. Dolan, "Emotion, cognition, and behavior," *Science*, vol. 298, no. 5596, pp. 1191–1194, 2002.

[100] C. G. Lange and W. James, *The Emotions*. Williams & Wikins, 1922.

[101] W. B. Cannon, "The James-Lange theory of emotions: a critical examination and an alternative theory," *The American Journal of Psychology*, vol. 39, no. 1, pp. 106–124, 1927.

[102] S. Schachter and J. E. Singer, "Cognitive, social, and physiological determinants of emotional state," *Physiological Review*, vol. 69, no. 5, pp. 379–399, 1962.

[103] S. N. Daimi and G. Saha, "Classification of emotions induced by music videos and correlation with participants' rating," *Expert Systems with Applications*, vol. 41, no. 13, pp. 6057–6065, 2014.

[104] X. Zhuang, V. Rozgic, and M. Crystal, "Compact unsupervised EEG response representation for emotion recognition," in *Proceedings of the IEEE International Conference on Biomedical and Health Informatics*, 2014, pp. 736–739.

[105] M. Chen, J. Han, L. Guo, J. Wang, and I. Patras, "Identifying valence and arousal levels via connectivity between EEG channels," in *Proceedings of the International Conference on Affective Computing and Intelligent Interaction*, 2015, pp. 63–69.

[106] ITU-T P.10, "Vocabulary for performance and quality of service," *International Telecommunication Union*, December 2006.

[107] K. Brunnstrom, S. A. Beker, K. D. Moor, A. Dooms, S. Egger, M.-N. Garcis, T. Hossfeld, S. Jumisky-Pyykko, C. Keimel, M.-c. Larabi, B. Lawlor, P. L. Callet, S. Moller, F. Pereira, M. Pereira, A. Perkis, J. Pibernik, A. Pinheiro, A. Raake, P. Reichl, U. Reiter, R. Schatz, P. Schelkens, L. Skorin-Kapov, D. Strohmeier, C. Timmerer, M. Varela, I. Wechsung, J. You, and A. Zgank, "Qualinet white paper on definitions of quality of experience," *Qualinet White Paper on Definitions of Quality of Experience Output from the 5th Qualinet Meeting*, 2013.

[108] X. K. Yang, W. S. Ling, Z. K. Lu, E. P. Ong, and S. S. Yao, "Just noticeable distortion model and its applications in video coding," *Signal Processing: Image Communication*, vol. 20, no. 7, pp. 662–680, 2005.

[109] A. Khan, L. Sun, and E. Ifeachor, "QoE prediction model and its application in video quality adaptation over UMTS networks," *IEEE Transactions on Multimedia*, vol. 14, no. 2, pp. 431–442, 2012.

[110] I. Kant, *Critique of Judgment*. Hackett Publishing, 1987.

[111] D. Joshi, R. Datta, E. Fedorovskaya, Q.-T. Luong, J. Z. Wang, J. Li, and J. Luo, "Aesthetics and emotions in images," *IEEE Signal Processing Magazine*, vol. 28, no. 5, pp. 94–115, 2011.

[112] K. Ukai and P. A. Howarth, "Visual fatigue caused by viewing stereoscopic motion images: background, theories, and observations," *Displays*, vol. 29, no. 2, pp. 106–116, 2008.

[113] Emotiv. [Online]. Available: https://emotiv.com/

[114] OpenBCI. [Online]. Available: http://openbci.com/

[115] NeuroSky EEG biosensor. [Online]. Available: http://neurosky.com/biosensors/eeg-sensor/

[116] Mitsar portable EEG system. [Online]. Available: http://www.mitsar-medical.com/eeg-system/portable-eeg/

[117] Avatar EEG. [Online]. Available: http://avatareeg.com/

[118] Melomind. [Online]. Available: http://www.melomind.com/

[119] K. B. Mikkelsen, S. L. Kappel, D. P. Mandic, and P. Kidmose, "EEG recorded from the ear: characterizing the Ear-EEG method," *Frontiers in Neuroscience*, vol. 9, pp. 438:1–8, 2015.

[120] S. Davis, E. Cheng, I. Burnett, and C. Ritz, "Multimedia user feedback based on augmenting user tags with EEG emotional states," in *Proceedings of the 3rd International Workshop on Quality of Multimedia Experience*, 2011, pp. 143–148.

[121] A.-N. Moldovan, I. Ghergulescu, S. Weibelzahl, and C. H. Muntean, "User-centered EEG-based multimedia quality assessment," in *Proceedings of the International Symposium on Broadband Multimedia Systems and Broadcasting*, 2013, pp. 1–8.

[122] Microsoft Band. [Online]. Available: https://www.microsoft.com/Microsoft-Band/en-us

[123] Jawbone UP. [Online]. Available: https://jawbone.com/up

[124] Basis Peak. [Online]. Available: https://www.mybasis.com/

[125] J. Frey, "Comparison of an open-hardware electroencephalography amplifier with medical grade device in brain-computer interface applications," in *Proceedings of the International Conference on Physiological Computing Systems*, 2016.

[126] H. P. Martinez, Y. Bengio, and G. N. Yannakakis, "Learning deep physiological models of affect," *IEEE Computational Intelligence Magazine*, vol. 8, no. 2, pp. 20–33, 2013.

[127] D. Wang and Y. Shang, "Modeling physiological data with deep belief networks," *International Journal of Information and Education Technology*, vol. 3, no. 5, pp. 505–511, 2013.

[128] X. Yang, T. Zhang, and C. Xu, "Cross-domain feature learning in multimedia," *IEEE Transactions on Multimedia*, vol. 17, no. 1, pp. 64–78, 2015.

[129] J. Ngiam, A. Khosla, M. Kim, J. Nam, H. Lee, and A. Y. Ng, "Multimodal deep learning," in *Proceedings of the 28th International Conference on Machine Learning*, 2011, pp. 689–696.

[130] D. E. Berlyne, *Aesthetics and Psychobiology*. Appleton-Century-Crofts, 1971.

[131] F. H. Farley and C. A. Weinstock, "Experimental aesthetics: children's complexity preference in original art and photoreproductions," *Bulletin of the Psychonomic Society*, vol. 15, no. 3, pp. 194–196, 1980.

[132] D. H. Saklofske, "Aesthetic complexity and exploratory behavior," *Perceptual and Motor Skills*, vol. 41, no. 2, pp. 363–368, 1975.

[133] C. Imamoglu, "Complexity, liking and familiarity: architecture and non-architecture Turkish students' assessments of traditional and modern house," *Journal of Environmental Psychology*, vol. 20, no. 1, pp. 5–16, 2000.

[134] Y. Gucluturk, R. H. A. H. Jacobs, and R. v. Lier, "Liking versus complexity: decomposing the inverted U-curve," *Frontiers in Human Neuroscience*, vol. 10, pp. 112:1–11, 2016.

[135] T. H. Falk, Y. Pomerantz, K. Laghari, S. Moller, and T. Chau, "Preliminary findings on image preference characterization based on neurophysiological signal analysis: toward objective QoE modeling," in *Proceedings of the 4th International Workshop on Quality of Multimedia Experience*, 2012, pp. 146–147.

[136] D. Wang, R. L. Buckner, M. D. Fox, D. J. Holt, A. J. Holmes, S. Stoeklein, G. Langs, R. Pan, T. Qian, K. Li, J. T. Baker, S. M. Stufflebeam, K. Wang, X. Wang, B. Hong, and H. Liu, "Parcellating cortical functional networks in individuals," *Nature Neuroscience*, vol. 18, no. 2, pp. 1853–1860, 2015.

[137] N. Jatupaiboon, S. Pan-ngum, and P. Israsena, "Real-time EEG-based happiness detection system," *The Scientific World Journal, Article ID 618649*, 2013.

[138] T. H. Falk, M. Guirgis, S. Power, and T. T. Chau, "Taking NIRS-BCIs outside the lab: towards achieving robustness against environment noise," *IEEE Transactions on Neural Systems and Rehabilitation Engineering*, vol. 19, no. 2, pp. 136–146, 2011.

[139] V. Menkovski, A. Oredope, A. Liotta, and A. C. Sanchez, "Predicting quality of experience in multimedia streaming," in *Proceedings of the 7th International Conference on Advanced in Mobile Computing and Multimedia*, 2009, pp. 52–59.

[140] W. Song, D. Tjondronegoro, and I. Himawan, "Acceptability-based QoE management for user-centric mobile video delivery: a field study evaluation," in *Proceedings of the ACM International Conference on Multimedia*, 2014, pp. 267–276.

[141] R. Gupta, H. J. Banville, and T. H. Falk, "PhySyQX: a database for physiological evaluation of synthesised speech quality-of-experience," in *Proceedings of the IEEE Workshop on Applications of Signal Processing to Audio and Acoustics*, 2015, pp. 1–5.

[142] A. Savran, K. Ciftci, G. Chanel, J. C. Mota, L. H. Viet, B. Sankur, L. Akarun, A. Caplier, and M. Rombaut, "Emotion detection in the loop from brain signals and facial images," in *Proceedings of the 2nd Summer Workshop on Multimodal Interfaces*, 2006.

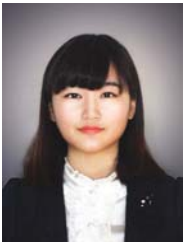

**Seong-Eun Moon** received her B.S. degree in mechanical engineering from Chiba University, Japan, in 2013. She is now with the School of Integrated Technology of Yonsei University and is working toward the Ph.D. degree. Her research interests include multimedia signal processing and physiological signal processing.

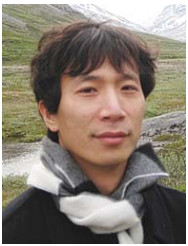

**Jong-Seok Lee** (M'06-SM'14) received his Ph.D. degree in electrical engineering and computer science in 2006 from KAIST, Korea, where he also worked as a postdoctoral researcher and an adjunct professor. From 2008 to 2011, he worked as a research scientist at Swiss Federal Institute of Technology in Lausanne (EPFL), Switzerland. Currently, he is an associate professor in the School of Integrated Technology at Yonsei University, Korea. His research interests include multimedia signal processing and machine learning. He is an author or co-author of over 100 publications. He serves as an Area Editor for the Signal Processing: Image Communication.